\documentclass[10pt,journal,compsoc]{IEEEtran}

\ifCLASSOPTIONcompsoc
  \usepackage[nocompress]{cite}
\else
  \usepackage{cite}
\fi

\usepackage{lipsum,url,booktabs,tabularx,multirow}
\usepackage[pdftex]{graphicx}
\usepackage{amsmath,amssymb}
\usepackage{chngcntr}
\DeclareMathOperator*{\argmax}{argmax}
\usepackage[figuresright]{rotating}
\makeatletter
\let\MYcaption\@makecaption
\makeatother

\usepackage[font=footnotesize]{subcaption}

\makeatletter
\let\@makecaption\MYcaption
\makeatother

\usepackage{array}
\usepackage{fixltx2e}

\usepackage{stfloats}
\begin{document}
%
\title{Cutting the Unnecessary Long Tail: Cost-Effective Big Data Clustering in the Cloud}
%
%
%
%

\author{Dongwei~Li,
        Shuliang~Wang,
        Nan~Gao,
        Qiang~He,
        and~Yun~Yang
\IEEEcompsocitemizethanks{\IEEEcompsocthanksitem D. Li is with the School of Computer Science, Beijing Institute of Technology, Beijing, China, and the School of Software and Electrical Engineering, Swinburne University of Technology, Victoria, Australia.\protect\\
E-mail: dongwei@bit.edu.cn, dli@swin.edu.au
\IEEEcompsocthanksitem S. Wang is with the School of Computer Science, Beijing Institute of Technology, Beijing, China.

E-mail: slwang2011@bit.edu.cn
\IEEEcompsocthanksitem N. Gao is with the School of Science, Royal Melbourne Institute of Technology, Victoria, Australia.

E-mail: nan.gao@rmit.edu.au

\IEEEcompsocthanksitem 
Q. He and Y. Yang are with the School of Software and Electrical Engineering, Swinburne University of Technology, Victoria, Australia.

E-mail: {qhe, yyang}@swin.edu.au
}
\thanks{}}

%
%

\markboth{}%
{Shell \MakeLowercase{\textit{et al.}}: Bare Demo of IEEEtran.cls for Computer Society Journals}
%



\IEEEtitleabstractindextext{%
\begin{abstract}
Clustering big data often requires tremendous computational resources where cloud computing is undoubtedly one of the promising solutions. However, the computation cost in the cloud can be unexpectedly high if it cannot be managed properly. The long tail phenomenon has been observed widely in the big data clustering area, which indicates that the majority of time is often consumed in the middle to late stages in the clustering process. In this research, we try to cut the unnecessary long tail in the clustering process to achieve a sufficiently satisfactory accuracy at the lowest possible computation cost. A novel approach is proposed to achieve cost-effective big data clustering in the cloud. By training the regression model with the sampling data, we can make widely used k-means and EM (Expectation-Maximization) algorithms stop automatically at an early point when the desired accuracy is obtained. Experiments are conducted on four popular data sets and the results demonstrate that both k-means and EM algorithms can achieve high cost-effectiveness in the cloud with our proposed approach. For example, in the case studies with the much more efficient k-means algorithm, we find that achieving a 99\% accuracy needs only 47.71\%-71.14\% of the computation cost required for achieving a 100\% accuracy while the less efficient EM algorithm needs 16.69\%-32.04\% of the computation cost. To put that into perspective, in the United States land use classification example, our approach can save up to \$94,687.49 for the government in each use.
\end{abstract}

\begin{IEEEkeywords}
Cloud computing, cost-effectiveness, clustering algorithms, big data, data mining.
\end{IEEEkeywords}}

\maketitle

\IEEEdisplaynontitleabstractindextext

%
\IEEEpeerreviewmaketitle

\IEEEraisesectionheading{\section{Introduction}\label{sec:introduction}}

%
%
%
%
\IEEEPARstart{T}{he} long tail refers to the phenomenon where the portion of the distribution has a large number of occurrences far from the ‘head’ or central part of the distribution, which is commonly observed in recommendation systems and data mining \cite{park2008long,chen2016long}. In recent years, with the explosive growth of data in many areas such as remote sensing \cite{zhang2016kira,ma2015remote}, business \cite{wixom2014current}, and bioinformatics \cite{alyass2015big}, the capability for data generation becomes so powerful and enormous. Clustering algorithms have been widely used as one of the most powerful meta-learning tools for accurate analysis of massive volumes of data generated by modern devices. The main goal of clustering is to categorize data points into clusters such that those grouped in the same cluster are similar according to specific metrics. During the clustering process, it is usual that the clusters are formed quickly at the early stage while changes slowly during the middle to late stages. This is the ‘long tail’ in clustering \cite{he2017cost}.

In the area of clustering, there have been lots of attempts to analyze and categorize the data for the huge number of applications. However, one of the major issues in using clustering algorithms is that it often requires tremendous computational resources especially when processing large-scale data sets. To illustrate this, we use the k-means algorithm to cluster remote sensing images. For $k$  clusters and $p$ pixels, a total of  $k \times p$ distances need be computed at each iteration. For example, for 10 classes and 40000 ($200 \times 200$) pixels, 50 iterations of the k-means clustering require 20 million multiplications for every image. Usually, the remote sensing data sets are huge and consist of tens of thousands of images such as SAT-6 \cite{basu2015deepsat}, AID \cite{xia2017aid}, NWPU-RESISC45 \cite{cheng2017remote}. As a result, processing such data is undoubtedly computationally intensive and extremely costly.

Small and medium-sized organizations usually cannot afford the exorbitant in-house IT infrastructure for processing such a large amount of data. Naturally, cloud computing, the latest distributed computing paradigm which eliminates the need to maintain expensive computing hardware, dedicated space, and software, becomes the best choice for them \cite{shen2017performance}.

Cloud computing adopts the pay-as-you-go model, where users are charged flexibly according to the usage of cloud services such as computational resources. However, the computation cost in the cloud can be unexpectedly high if users cannot manage it properly, which also becomes a bottleneck for big data mining in the cloud. For instance, running 50 m4-2xlarge EC2 virtual machine (VM) instances in Amazon’s Sydney datacenter costs \$18,000 per month \cite{Amazon}.

In most clustering situations, it is not always necessary to achieve the optimal solution because users often do not need 100\%. Take the marketing for example, based on various customer interests, age and product holding information, clustering techniques have been used for creating customer groups. In this situation, a reasonable margin of inaccuracy is acceptable because marketers do not need their customers to be grouped with 100\% accuracy. As long as they have a general picture of the clustering result, they are able to make a decision. In fact, there will never be completely accurate, e.g., weather forecasting or land use statistics. In such scenarios, stopping the clustering process at a reasonable point is important in saving computation costs if it is more preferable to achieve a sufficiently satisfied accuracy at a low computation cost than a 100\% accuracy at a high cost.

Thus, cutting the unnecessary ‘long tail’ in the clustering process is a promising solution to cost-effective clustering. In other words, we need to study how to achieve a sufficiently satisfactory clustering accuracy at the lowest possible computation cost.

Cost-effective clustering in the cloud allows big data analytics to be applied in a broader range of fields by more businesses and organizations, especially small and medium-sized ones with the limited budget. He et al. observed the long tail phenomenon and studied the cost effectiveness of the k-means algorithm in the cloud. They found that achieving 99\% accuracy with the k-means algorithm only needs a bit more than 20\% of computation time on average \cite{he2017cost}. However, when to stop the k-means algorithm automatically with the desired accuracy has not been well investigated by researchers up to now.

There is a variety of clustering techniques that can be adopted for exploration and demonstration of the cost effectiveness of big data clustering in the cloud. Among the top 10 data mining algorithms discussed by Wu et al. \cite{wu2008top}, k-means \cite{macqueen1967some} and EM (Expectation Maximization) algorithms \cite{dempster1977maximum} belong to the field of clustering. Furthermore, k-means and EM are both iterative algorithms and converge to the final (optimal) result iteratively \cite{selim1984k,wu1983convergence}, which provides possibilities for us to calculate the accuracy of the intermediate clustering result at each iteration of the clustering process. Therefore, we choose k-means and EM algorithms to explore and demonstrate the cost-effective clustering in the cloud.

The contributions of the paper are as follows:
\begin{enumerate}
    \item We demonstrated the ‘long tail’ phenomenon in the clustering process, and defined the cost effectiveness problem of k-means and EM clustering algorithms in the cloud.
    \item To the best of our knowledge, this is the first paper to achieve cost-effective clustering in the cloud through cutting the unnecessary ‘long tail’. We proposed a regression model between the change rate of objective function and clustering accuracy.
    \item We compared the excellent performance of cost effectiveness of k-means and EM algorithms on multiple benchmark data sets, and discussed the threats to validity of the results.
\end{enumerate}
The remainder of the paper is organized as follows. Section 2 presents a motivating example and analyzes the research problem. Then, Section 3 describes the methodologies used in the cost-effectiveness problem and Section 4 proposes a novel approach for cost-effective big data clustering in the cloud. Section 5 displays the results of experiments conducted on different data sets. Section 6 surveys the related work. Finally, Section 7 addresses the conclusions and future work.

\section{Motivating Example and Problem Analysis}
In this section, we introduce an example to motivate cost-effective big data clustering and then analyze the research problem.

\subsection{Motivating Example}
Knowledge about land use and land cover has become increasingly important in overcoming the problems such as uncontrolled development, deteriorating environmental quality, loss of prime agricultural lands, destruction of important wetlands, and loss of fish and wildlife habitat \cite{anderson1976land}. The U.S. Department of Agriculture reported that, during the 1960’s, a total of 730,000 acres were urbanized each year, transportation land uses expanded by 130,000 acres per year, and recreational area increased by about 1 million acres per year. The present distribution and area of agricultural, recreational, urban lands, as well as their changing proportions, are needed by legislators, planners, state and local government officials to determine better land use policies and implement effective plans for regional development. 

The recent advances in remote sensing techniques give birth to explosive growth of remote sensing images, which can be used effectively to calculate the current use of land sources. Generally, remote sensing images in the specified district have similar spectral characteristics and contain similar components such as “forest”, ”water”, “road”, “building”, “grassland” and “wasteland”. By clustering the pixels in remote sensing images that are spectrally similar, we can get an intuitive overview of remote sensing objects without any prior knowledge, which is significant in the classification statistics on land use. 

Suppose that the state governor plans to have statistics about the land use classification of California. According to the high solution aerial images from USGS National Map Urban Area Imagery collection, the partitioned remote sensing images should be extracted from the original data set, the solution of which is 1 foot per pixel. Since the California area is about 423,970 $km^2$, it is required to process tens of thousands of remote sensing images. The computation cost would be extremely high.
\label{}

\subsection{Problem Analysis}
Remote sensing clustering for land use classification is both computation- and data-intensive. For a single machine, the limitation of its hardware resources results in a bottleneck in processing such huge data. This bottleneck can be avoided if we run the remote sensing images clustering in the cloud.
The cloud can offer virtually unlimited computational resources for processing large data sets. Cloud computing adopts the “pay-as-you-go” model \cite{armbrust2010view} and enables flexible and on-demand access to computational resources, which allow big data clustering to be performed by using only necessary computational resources for a needed period of time. Since the cloud cost is the main concern for users, how to achieve cost effectiveness has become a critical issue for both academia and industry.
For clustering applications such as land use classification, it is usually acceptable for the governor if the accuracy is within a reasonable range. Due to the long tail phenomenon (see Section 3.4) in the clustering process, a sufficient clustering accuracy may be obtained within a short time. After that, incremental accuracy improvement usually takes a relatively long time in the remainder of the clustering process. Thus, we need to consider the utilization of this phenomenon and find an appropriate point to terminate the clustering process to achieve satisfactory accuracy at a low cost.

\section{Methodology}
This section presents our study of cost-effective clustering, including the candidate clustering techniques, the accuracy calculation method, the cloud cost computing model, and cost-effective clustering analysis.
Clustering is a powerful method for analyzing massive volumes of data. The main idea of clustering is to minimize a certain criterion function usually taken up as a function of the deviations among all patterns from their respective cluster centers. Usually, the minimization of the criterion function is sought to utilize an iterative scheme that starts with a chosen initial cluster configuration of the data, then alters the cluster membership in an iterative manner to obtain a better configuration.
Appendix A lists the key notations used in this paper.

\subsection{Candidate Clustering Techniques}
Clustering is an unsupervised method for finding patterns based on features \cite{jain1999data}. Usually, a feature point can be represented by a vector $x=(x_1, x_2, ...,x_d)$. Based on the distance measure among feature vectors, a label will be assigned to each feature. Here, we take the popular k-means and EM algorithms as examples to demonstrate the cost effectiveness of big data clustering in the cloud.

\subsubsection{K-means Algorithm}
The k-means algorithm proposed by Mac Queen \cite{macqueen1967some} is one of the simplest and most popular techniques in data mining. It begins with $k$ initial centers and each point will be assigned with a label based on the distance between the point and the cluster centers. The steps of the k-means algorithm are as follows:

Step 1: Select $k$ points as initial centers $C=\{c_1,c_2,...,c_k \}$.

Step 2: For each $i\in\{1,2,...,k\}$, set cluster $C_i$  as the set of data points that are closer to $c_i$ than to $c_j$ for all $j \neq i$. 

Step 3: Recompute $c_i$ as the center of $C_i$:
\begin{equation}
    c_i = \frac{1}{\left | C_i \right |} \sum_{x\in C_i}x.
\end{equation}

Step 4: Repeat Steps 2 and 3 until $C$ no longer changes.
During the process, let $\mu_i$ represent the mean of cluster $C_i$. Then the goal of k-means is to minimize the criterion function in an iterative manner:
\begin{equation}
    J = \sum_{i=1}^{k}\sum_{x\in C_i}{\left\|x-\mu_i\right\|}^2
    \end{equation}

In Equation (2), the squared Euclidean distance is adopted to represent the metric of ${\left\|x_i-\mu_k \right\|}^2$ due to its computational simplicity since the cluster at each iteration can be calculated in a straightforward manner. 
The time complexity of the k-means algorithm is $O(nkdi)$, where $n$ is the number of $d$ dimensional data points in the data set, $k$ is the number of clusters and $i$ is the number of iterations for the clustering process to complete (i.e. converge).

\subsubsection{EM Algorithm}
The Expectation-Maximization (EM) algorithm is designed to estimate the maximum likelihood parameters of a statistical model in many situations, such as the one where the equations cannot be solved. EM approximates the unknown model parameters iteratively with the Expectation step ($E$ step) and the Maximization step ($M$ step) which are as follows:

\textbf{E step} calculates the expected value of the log-likelihood function, with respect to the conditional distribution of $Z$ given $X$ under the current estimate of the parameters $\theta^t$
\begin{equation}
    Q(\theta|\theta^t) = E_{Z| X,\theta^t}(\log L(\theta;X,Z)) 
\end{equation}

\textbf{M step} finds the parameters that maximize this quantity:
\begin{equation}
    \theta_{t+1} = \argmax_\theta Q(\theta|\theta^t) 
\end{equation}
The EM algorithm seeks to find the maximum likelihood estimation (MLE) by iterating the above two steps.
\subsection{Accuracy Calculation}
Accuracy is a crucial measurement for evaluating the effectiveness of big data clustering. For the purpose of demonstrating the gradual increase of the clustering accuracy iteration by iteration, we use the final clustering result as the reference partition noted by $P_f$ as 100\% accuracy. Through the comparison between the clustering results achieved at each iteration of the algorithm, we can demonstrate how the accuracy of the intermediate partition result $P_i \in \{P_1,P_2,...,P_f \}$ increases.

The accuracy can be measured by the similarity between $P_i$ and $P_f$. In our research, we use the \textit{Rand Index} \cite{rand1971objective} to assess the similarity, which is a popular accuracy calculation method in the data clustering field. The Rand Index measures the similarity between two data clustering partitions. Each partition is viewed as a collection of $n \times (n-1)/{2}$ pairs of elements, where $n$ is the size of the data set. For each pair of data points, a partition either assigns them to the same cluster or different clusters. Thus, the similarity between partitions $P_1$ and $P_2$ can be calculated as follows:
\begin{equation}
    Rand(P_1,P_2) = \frac{n_{11}+n_{00}}{n_{00}+n_{01}+n_{10}+n_{11}} = \frac{n_{11}+n_{00}}{\binom{n}{2}}
\end{equation}
where:

$n_{11}$: the number of pairs of elements that are placed in the same clusters both in $P_1$ and $P_2$;

$n_{00}$: the number of pairs of elements that are placed in the different clusters both in $P_1$ and $P_2$;

$n_{01}$: the number of pairs of elements that are placed in the same clusters in $P_1$ but in different clusters in $P_2$;

$n_{10}$: the number of pairs of elements that are placed in different clusters in $P_1$, but in the same clusters in $P_2$.

\begin{figure}
    \centering
    \includegraphics[width=0.9\linewidth]{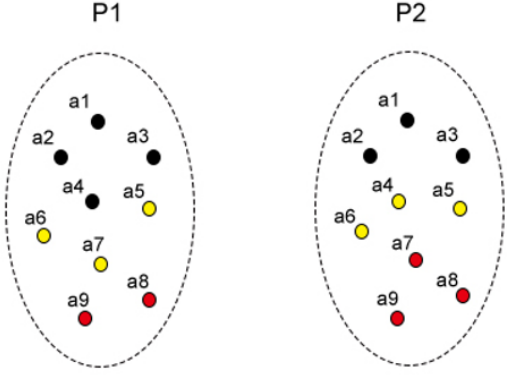}
    \caption{An example of calculating Rand Index between $P_1$ and $P_2$}
    \label{randindex}
\end{figure}

Using the Rand Index as the similarity calculation measure, we can compute the clustering accuracy at each iteration of the clustering process. Take Fig.~\ref{randindex} for example, for the pairs which are placed in the same cluster (i.e., same color) in $P_1$ and $P_2$ contains $(a1, a2)$, $(a1, a3)$, $(a2, a3)$, $(a5, a6)$, $(a8, a9)$. The pairs that are placed in different clusters in both $P_1$ and $P_2$ include $(a1, a5)$, $(a1, a6)$, $(a1, a7)$, $(a1, a8)$, $(a1, a9)$, $(a2, a5)$, $(a2, a6)$,$(a2, a7)$, $(a2, a8)$, $(a2, a9)$, $(a3, a5)$, $(a3, a6)$, $(a3, a7)$, $(a3, a8)$, $(a3, a9)$, $(a4, a7)$, $(a4, a8)$, $(a4, a9)$, $(a5, a8)$, $(a5, a9)$, $(a6, a8)$, $(a6, a9)$. Then, there is $Rand(P_1,P_2 )=(5+22)/36=75\%$. Obviously, the value of Rand Index increases with iterations and at the final iteration of clustering process, where $P_i=P_f$, there is $Rand (P_i,P_f )=1$, which indicates that the process completes with a 100\% accuracy.
\subsection{Cloud Computing Cost Model}
The cost of computation resources when clustering big data can be calculated by the cost models offered by cloud vendors. In this research, we use Amazon’s Elastic Compute Cloud (Amazon EC2) web services, which offer four different cost models: on-demand, reserved instances, spot instances, and dedicated hosts. The on-demand cost model is the basic cost model, under which computing capacities are paid for by the hours without long-term commitments or upfront payments. 

In this research, the on-demand cost model is employed to calculate the computation cost incurred during the clustering process:
\begin{equation}
    {Cost}_{comp} = {Price}_{unit} \times {Time}_{comp}
\end{equation}
Computation time $Time_{comp}$ is measured by the time taken by the clustering process. The unit price $Price_{unit}$ is decided by the computational resource employed in running the algorithm. Take EC2 for example, there are seven major categories of EC2 VM instances: Linux, SLES, RHEL, windows, windows with SQL Standard, Windows with SQL Web and Windows with SQL Enterprise. In different categories, there are various types of EC2 VM instances available at different unit prices. For instance, in Windows category, 36 EC2 instances are displayed for 4 types: General Purpose, Compute Optimized, Memory Optimized, and Storage Optimized. The unit prices differ across different areas and range from \$0.0066 to \$38.054 per hour.

In this research, we use the computation time as an indicator of the computation cost for simplicity. When we use a specific Amazon EC2 VM instance, it can be found that the computation cost and computation time are positively correlated. Generally, the longer the computation time, the higher the computation cost is.

Before running the algorithms, some other costs may occur such as the transfer cost and storage cost of the big data set in the cloud. However, the costs incurred by data storage and data transfer are independent of the clustering process. Thus, in this research, we focus only on the cost incurred by the computation of the clustering process and isolate it from the other costs.

\subsection{Cost-effective Clustering Analysis}
He et al. \cite{he2017cost} demonstrated the long tail phenomenon using the k-means algorithm as an example. The same long tail phenomenon can also be found in our experiments using both k-means and EM algorithms (see Section 5), which makes it possible to compute and demonstrate the accuracy of the intermediate clustering result with incurred cost at each iteration of the clustering process.
  
\begin{figure}
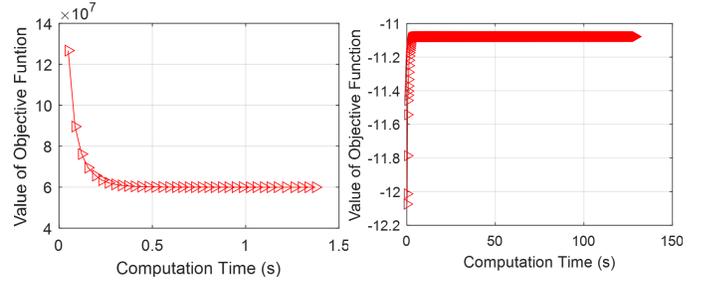

\begin{subfigure}{.5\linewidth}
    \centering
    \includegraphics[width=\linewidth]{image/objective1.pdf}
    \caption{The k-means algorithm}
    \label{fig:sfig1}
\end{subfigure}%
\begin{subfigure}{.5\linewidth}
    \centering
    \includegraphics[width=\linewidth]{image/objective2.pdf}
    \caption{The EM algorithm}
    \label{fig:sfig2}
\end{subfigure}
\caption{Objective function over computation time}
\label{objective}
\end{figure}

In the clustering process, the long tail phenomenon is based on the convergence property of clustering algorithms. In the k-means algorithm, the objective function (sum of mean square of all points) is monotonically decreasing iteratively and can converge in finite steps. A rigorous proof of convergence property for k-means is given in \cite{selim1984k}. For the EM algorithm, the objective function (log likelihood) is monotonically increasing and guaranteed to find a local maximum for the model parameters estimate \cite{wu1983convergence}. From Fig.~\ref{objective}, we can see the change in the value of the objective function over computation time.

\begin{figure}
    \centering
    \includegraphics[width=0.85\linewidth]{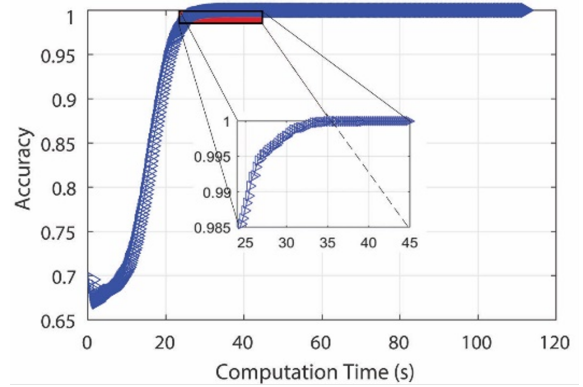}
    \caption{The long tail phenomenon in the clustering process}
    \label{longtail}
\end{figure}
As the clustering process continues and the clustering results stabilizes gradually, the convergence of k-means and EM become very slow, especially at the middle to late stage of clustering, which incur high costs for big data clustering. Fig.~\ref{longtail} shows the long tail phenomenon in the clustering process, where y-axis means the clustering accuracy calculated using Rand Index.

For the cost-effective clustering problem, the convergence rates of the objective functions for k-means and EM at a certain point can be analyzed with the clustering accuracy at the same time. This way, we can explore the relationship between them and propose a solution to cost-effective big data clustering in the cloud.

\section{Proposed Approach for Big Data Clustering in the Cloud}

In the k-means and EM clustering process, the objective function $J$ monotonically changes iteratively until converges in finite steps. For $i\in\{1,2,...,f\}$, the clustering partition $P_i$ is updated one iteration after another and gradually approaches the final result $P_f$, which means that the label of each point is updated iteratively and clustering accuracy $Rand(P_i,P_f)$ approaches 100\%. Therefore, we set $r_i= Rand(P_i,P_f)$ to represent the clustering accuracy at the $i$th iteration.

\begin{figure}
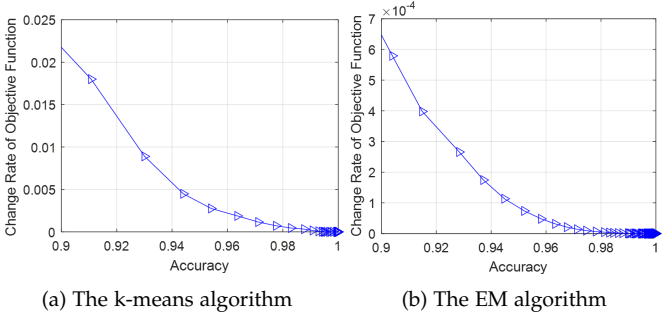

\begin{subfigure}{0.496\linewidth}
    \centering
    \includegraphics[width=1.04\linewidth]{image/rate1.pdf}
    \caption{The k-means algorithm}
    \label{fig:sfig3}
\end{subfigure}
\begin{subfigure}{0.496\linewidth}
    \centering
    \includegraphics[width=0.96\linewidth]{image/rate2.pdf}
    \caption{The EM algorithm}
    \label{fig:sfig4}
\end{subfigure}
\caption{The change rate of objective function over accuracy}
\label{rate}
\end{figure}

The objective functions of k-means and EM algorithms are both monotonic and tend to converge over iterations \cite{selim1984k,wu1983convergence}. However, the value of objective function can be extremely dissimilar for different clustering algorithms (see Fig.~\ref{objective}) and can not be compared directly through the single value. Even with the same algorithm, the different distribution of data will lead to very distinct value of objective function. Therefore, in this research, we define the change rate of the value of the objective function at the $i$th iteration of the clustering process by $h_i$:  

\begin{equation}
    h_i = \frac{|J_i-J_{i-1}|}{|J_{i-1}|}, i\in \{2,3,...f\}
\end{equation}
where $J_i$ indicates the value of the objective function at the $i$th iteration during the clustering process. As the clustering converges, the accuracy $r_i$ increases to 1 while $h_i$ decreases to 0. Therefore, there is a significant negative correlation between $h_i$ and $r_i$ (see Fig.~\ref{rate}).

In big data clustering, given a set of data points $D$, a random sampling strategy needs to be adopted first, and then the data set is partitioned into $n$ groups and each group has $k=D/n$ individuals. Random sampling is also called probability sampling, where each subject of $k$ individuals has the same probability of being chosen for the samples as other subjects of $k$ individuals \cite{yates2007practice}. Therefore, when  the random sampling is done, each sample is an unbiased representation of the entire data and has the very same distribution pattern as other samples. After that, $n$ samples are split into training set and validation set. 

In the training set, we utilize the regression analysis to develop a prediction model for estimating the relationship between $h_i$ and $r_i$ during the clustering process. The first concern is how to select the best regression model. In statistics, the sum squares due to error (SSE), R-square, adjusted R-square, and root mean squared error (RMSE) are commonly used as standard statistical metrics \cite{draper2014applied} for measuring the performance of the regression model. Generally, the closer the SSE and RMSE are to 0, the better the regression model selection and fitting, hence the more successful the data forecast. R-square and adjusted R-square range between 0 and 1, with a value closer to 1 indicating a better fit.

Based on comprehensive experiments, we found that the quadratic polynomial regression model shows the best fit than other popular regression models in most cases, such as linear regression, three-degree polynomial regression, exponential regression, logistic regression, lasso regression, etc. The quadratic polynomial model is as follows \cite{myers1990classical}: 

\begin{equation}
    h_i = \beta_0 + \beta_1 \times r_i+\beta_2\times{r_i}^2+\varepsilon, i\in \{1,2,...f\},
\end{equation}
where $\varepsilon$ is an unobserved random error with mean zero conditioned on a scalar variable $Rand(i)$. $\beta_0$, $\beta_1$, $\beta_2$ are estimated parameters which represent the relationship between $h_i$ and $r_i$. 

By establishing the regression model between these two variables in the training data set, we can estimate the changes in $h_j$ against the changes in $r_j$. Then, we conduct the clustering process iteratively in the validation data set when $h_i \leq h_j$. After that, we can terminate the clustering process to reduces unnecessary iterations and save computation costs.

To evaluate the proposed approach, we define the total computation time first. The total computation time $\textit{Time}_\textit{comp}$ includes the overall clustering time for the training data set $\textit{Time}_\textit{train}$, and the early-stop computation time $\textit{Time}_\textit{actual}$ when clustering reaches the desired accuracy, which can be calculated as:

\begin{equation}
    \textit{Time}_\textit{comp} = \textit{Time}_\textit{train} + \textit{Time}_\textit{actual}
\end{equation}

The training process is conducted only once. When it is finished, the regression model can be applied repeatedly for many applications. Thus, $\textit{Time}_\textit{train}$ is negligible compared to the overall cost in the long term (see Section 5.4 for the corresponding experimental analysis). Since computation time is the only indicator of the cost in our research, the cost effectiveness percentage $\textit{Cost}_\textit{effective}$ can also be represented as follows:

\begin{equation}
    \textit{Cost}_\textit{effective} \approx \frac{\textit{Time}_\textit{actual}}{\textit{Time}_\textit{full}}
    \label{costeffective}
\end{equation}
where $\textit{Time}_\textit{full}$ means the expected computation time in the clustering when achieves a 100\% accuracy. The smaller the value of $\textit{Cost}_\textit{effective}$ is, the more cost effective the clustering will be.

\section{Experiment Result}
In this section, we first describe the data sets and the experimental settings. Then, we evaluate the cost-effectiveness of the proposed approach. Finally, we discuss the performance of different clustering algorithms and illustrate the threats to validity.
\subsection{Data Set Description}
We have applied our approach to the 3D Road Network, Skin Segmentation, Poker Hand data sets from UCI machine learning repository and the SpaceNet data set of high-solution satellite images from DigitalGlobe (see Table~\ref{datasets}). The above data sets are the benchmarks for many studies in machine learning research and have been cited in high-impact peer-reviewed venues \cite{chen2014efficient,di2013novel,bifet2009new,vakalopoulou2017integrating}.

\begin{table}[]
\caption{Description of the extracted features}\label{datasets}
\begin{tabular}{llll}
\hline \\[-1em]
\textbf{\textit{Dataset}}                    & \textbf{\textit{Instances}}                   & \textbf{\textit{Attributes}} &\textbf{\textit{Classes}} \\ \hline  \\[-1em]
\textit{3D Road Network}   & 434,874                     & 4          & 4, 8    \\ \\[-1em]
\textit{Skin Segmentation} & 245,057                     & 4          & 2       \\ \\[-1em]
\textit{Poker Hand}        & 1,025,010                   & 11         & 10      \\ \\[-1em]
\textit{SpaceNet}          & \textgreater{}3,117,858,324 & 3          & 6       \\ \hline \\[-1em]
\end{tabular}
\end{table}

The 3D Road Network data set has a total of 434,874 3-dimensional data points without class labels. It contains the longitude, latitude and altitude information about a road network covering a region of $185 \times 135$ ${km}^2$ in North Jutland, Denmark. 

The Skin Segmentation data set has a total of 245,057 instances and is collected by randomly sampling B, G, R values from face images of various age groups, race groups and genders obtained from FERET database and PAL database. The data set is made up of 2 classes: the skin samples and non-skin samples.

The Poker Hand data set consists of 1,025,010 records and each record is an example of a hand consisting of five playing cards drawn from a standard deck of 52. Each card is described with two attributes (suit and rank), for a total of 10 predictive attributes. There is one attribute (class) that describes the "Poker Hand".

The SpaceNet data set is an online repository of freely available satellite imagery collected from DigitalGlobe's commercial satellites that includes more than 17,533 high-resolution images ($438 \times 406$ pixels) in Rio De Janeiro, Las Vegas, Shanghai, and Khartoum areas. This data set contains a wealth of geospatial information relevant to many downstream use cases such as infrastructure mapping and land use classification. 

\subsection{Experimental Setup}
Given the data sets at hand, the main purpose of the experimental setup is to use a default configuration on the parameters of the clustering algorithms. In general, finding an optimal number of clusters is an ill-posed problem of crucial relevance in clusters analysis \cite{fahad2014survey}. Thus, we have chosen the number of clusters with respect to the number of unique class labels in the Skin Segmentation (2 classes) and the Poker Hand (10 classes) data sets. Since the 3D Road Network data set does not have class labels, we ran the data set with $k = 4, 8$. Usually, the number of clustering for remote sensing images is lower than 10 and can be set in required scenarios \cite{richards1999remote}. Thus, with the SpaceNet data set, we attempt to partition the images into six regions of pixels that can be given a common label, such as “forest”, ”water”, “road”, “building”, “grassland” and “wasteland” for the land use classification as described in the motivating example, i.e., $k = 6$.

For non-image data sets, including the Skin Segmentation, the Poker Hand and the 3D Road Network data sets, a random sampling generation strategy was applied. In our research, for data set consisted of $n$ points, we randomly select $m$ data with $n/m$ times. For example, for a data set consisted of 500,000 points, it can be divided into $20,000\times25$ (25 groups and each group has 20,000 points), $10,000\times50$, $5,000\times100$ etc. After extensive experiments, we found that when the data set has more groups and group size is larger (which means that we need to find a balance to make both of the group’s number and size not too small), the experimental result usually shows better performance. Generally, when each group's size is above 10,000 points and the number of groups is above 50, our approach achieves better results. The above phenomenon also indicates that the larger the data set, the more effective our method is. 
For SpaceNet imagery data set, since each satellite image has $438\times406$ data points, we regard each image as a sampling group for simplicity and there are 17,533 groups in total. Although the sampling size of SpaceNet data set is larger than the non-image data sets, this grouping strategy is still reasonable considering huge number of groups.

In the experiments, we use the 10-fold cross-validation to divide the groups into the training set and the validation set. For image data set (SpaceNet), each image is considered as a group. As the remote sensing data set is huge, we select 100 sample images as the training data set that can simulate the regression model quite accurately.

The experiments were implemented on Matlab r2013a and conducted on a machine with a 2.20 GHz Intel (R) Core (TM) i3 processor and 10G memory. The operating system is 64-bit Windows 7 enterprise.

\subsection{Experimental Performance}
In this section, we present and discuss the results achieved by the candidate clustering algorithms for the given data sets. Firstly, a data set sampling strategy is applied and the data sets are then divided into training data set and validation data set. We will introduce the experimental performance of our approach in the training process and validation process.

\subsubsection{Training Process}

\begin{figure}
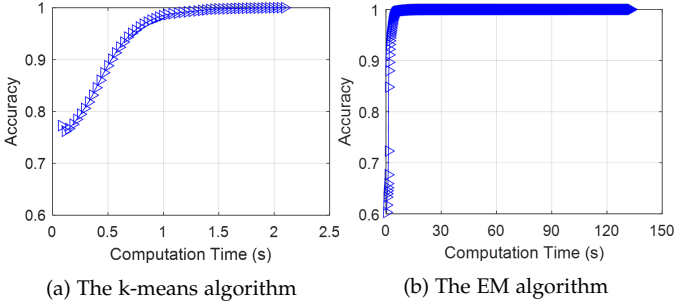

\begin{subfigure}{.5\linewidth}
    \centering
    \includegraphics[width=\linewidth]{image/accuracy1.pdf}
    \caption{The k-means algorithm}
    \label{fig:sfig5}
\end{subfigure}%
\begin{subfigure}{.5\linewidth}
    \centering
    \includegraphics[width=\linewidth]{image/accuracy2.pdf}
    \caption{The EM algorithm}
    \label{fig:sfig6}
\end{subfigure}
\caption{The clustering accuracy over computation time }
\label{accuracy}
\end{figure}

\textbf{Illustrating Long Tail Phenomenon}. Fig.~\ref{accuracy} shows the increase in the clustering accuracy over iterations during the clustering process for one group from the training set. Each marker on the curve indicates the intermediate partition at every iteration. It can be seen that the k-means algorithm first takes a relatively small number of iterations (19 iterations) to reach a high accuracy (95.06\%), and then takes a large number of iterations (37 more iterations) to converge to the accuracy of 100\%. This confirms the long tail phenomenon discussed in Section 3.3, which indicates that the majority of computation time is consumed at the middle to late stages. In our experiments, we also observed the long tail phenomenon with the different data sets using both k-means and EM, which indicates the feasibility of stopping at an early point of the clustering to achieve the desired accuracy.

In addition, fluctuations of accuracy may be observed in the early stage in clustering which is normal due to the chosen initial points. However, how to select the optimal initial points is not part of this research, so we do not discuss it in this paper.

\begin{figure}
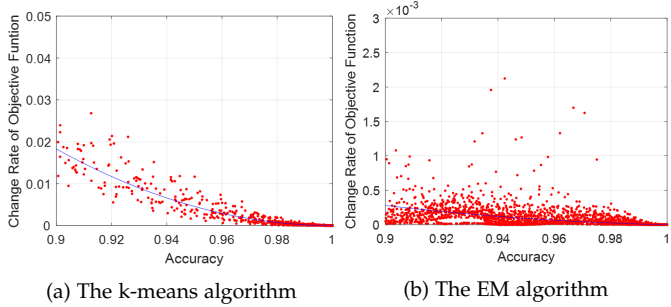

\begin{subfigure}{.5\linewidth}
    \centering
    \includegraphics[width=\linewidth]{image/changerate1.pdf}
    \caption{The k-means algorithm}
    \label{fig:sfig7}
\end{subfigure}%
\begin{subfigure}{.5\linewidth}
    \centering
    \includegraphics[width=\linewidth]{image/changerate2.pdf}
    \caption{The EM algorithm}
    \label{fig:sfig8}
\end{subfigure}
\caption{The regression model in training set (3D Network Road k=4)}
\label{changerate}
\end{figure}

\textbf{Building Regression Model}. We have explored the relationship between the change rate of the value of objective function $h_i$ calculated using (8) and clustering accuracy $r_i$ at the same iteration. In Fig.~\ref{fig:sfig7}, we see the relationships between $h_i$ and $r_i$ from all groups in the training data set (3D Road Network k = 4) by the k-means algorithm, which is represented by a series of scattered points. Then we can obtain the regression model by Matlab cftool box \cite{matlabcurve} through the points using (8) as follows:
$$h_i = 1.83 \times {r_i}^2-3.66 \times r_i+1.83$$
This regression model illustrates the general relationship between $h_i$ and $r_i$ in the k-means algorithm. Similarly, we obtain the regression model by the EM algorithm using (8) as follows (see Fig.~\ref{fig:sfig8}):
$$h_i=0.007232 \times {r_i}^2-0.01479 \times r_i+0.007558 $$
\textbf{Setting Desired Accuracy}. Then, we can set the desired accuracy $r_j$ and calculate the corresponding $h_j$ through the regression model obtained from the training process. Here, due to the page limit, we only consider the situations when desired accuracies are set for $r_j$ = 90\%, 95\%, 99\% and 99.9\% which we believe are sufficient. Table~\ref{relation} displays the relationship between the desired accuracy and change rate of the value of the objective function intuitively in 3D Road Network data set when $k=4$. 

\begin{table}[]
\caption{The relation between accuracy and change rate of objective functions in K-means and EM}
\label{relation}
\centering
\begin{tabular}{@{}lllll@{}}
\toprule
\begin{tabular}[c]{@{}l@{}} \textbf{Desired Accuracy}\end{tabular} & \textbf{90\%}    & \textbf{95\%}    & \textbf{99\%}    & \textbf{99.9\%}  \\ \midrule
\textit{  $h_j$ (k-means)}                                     & 1.83e-2 & 4.60e-3 & 1.83e-4 & 1.83e-6 \\\\[-1em]
\textit{  $h_j$ (EM)}                                          & 1.05e-4 & 3.44e-5 & 3.98e-6 & 3.33e-7 \\ \bottomrule
\end{tabular}
\end{table}

The candidate clustering algorithms are terminated in the iterative process once the change rate of the objective function $h_i$ is below the set value $h_j$, i.e., $h_i\leq h_j$. In a real-world application, the clustering task will stop when it reaches the desired accuracy. 

\subsubsection{Validation Process}
To validate the performance of the proposed approach, we mainly focus on two aspects: cost effectiveness and achieved accuracy.

\textbf{Cost-effectiveness Validation}. We run validation set for different data sets and obtain the total clustering completion time $Time_{full}$ for k-means and EM algorithms. Fig.~\ref{chart} shows the percentages of actual computation time $Time_{actual}$ by using our approach in different data sets. By setting the desired accuracy, the candidate algorithms can stop at an early point. After using Eq.~(\ref{costeffective}), the average actual computation time is only at 23.74\%, 33.50\%, 56.57\% and 81.06\% of the total time when desired accuracies are 90\%, 95\%, 99\%, and 99.9\% respectively using the k-means algorithm. Similarly, for EM algorithm, the average actual computation time accounts for 9.4\%, 14.46\%, 20.73\%, and 32.78\% respectively of the total computation time. Since the cloud computation cost is directly related to computation time, both k-means and EM algorithms can achieve high cost effectiveness in the clustering process in the cloud using our approach.
\begin{table*}[t]
\caption{Achieved accuracy for K-means}
\label{kmeans}
\centering
\begin{tabularx}{\textwidth}{XXXXX}
\hline \\[-0.8em]
\multirow{2}{*}{\textbf{Dataset / k}} & \multicolumn{4}{l}{\textbf{Desired Accuracy (Standard Deviation)}}                           \\\\[-1em] \cline{2-5} \\[-0.8em]
                             & \textgreater{}\textbf{90\%} & \textgreater{}\textbf{95\%} & \textgreater{}\textbf{99\%} & \textgreater{}\textbf{99.9\%} \\ \hline \\[-0.8em]
3D\_Road / 4                 & 91.67\% (0.1670)   & 95.84\% (0.0094)   & 99.14\% (0.0040)   & 99.93\% (0.0015)     \\\\[-1em]
3D\_Road / 8                 & 90.92\% (0.0207)   & 94.33\% (0.0278)   & 97.78\% (0.0226)   & 99.77\% (0.0059)     \\\\[-1em]
Skin\_Seg / 2                & 91.79\% (0.0068)   & 96.86\% (0.0050)   & 98.87\% (0.0036)   & 99.75\% (0.0013)     \\\\[-1em]
\textit{Poker\_Hand / 10}    & 94.00\% (0.0235)   & 95.58\% (0.0240)   & 98.09\% (0.0211)   & 99.80\% (0.0061)     \\\\[-1em]
\textit{SpaceNet / 6}        & 92.05\% (0.0349)   & 94.75\% (0.0341)   & 98.64\% (0.0098)   & 99.79\% (0.0036)     \\ \hline\\[-0.8em]
Average                      & 92.09\%            & 95.47\%            & 98.50\%            & 99.81\%              \\ \hline\\[-1em]
\end{tabularx}

\end{table*}
\begin{table*}[t] 
\caption{Acheived accuracy for EM}
\label{em}
\centering
\begin{tabularx}{\textwidth}{XXXXX}
\hline \\[-0.8em]
\multirow{2}{*}{\textbf{Dataset / k}} & \multicolumn{4}{l}{\textbf{Desired Accuracy (Standard Deviation)}}                           \\\\[-1em] \cline{2-5} \\[-0.8em]
                             & \textgreater{}\textbf{90\%} & \textgreater{}\textbf{95\%} & \textgreater{}\textbf{99\%} & \textgreater{}\textbf{99.9\%} \\ \hline \\[-0.8em]
3D\_Road / 4                 & 90.71\% (0.1599)   & 95.16\%  (0.0551)  &98.07\% (0.0310)  & 99.63\% (0.0015)     \\\\[-1em]
3D\_Road / 8                 & 91.67\% (0.0467)   & 95.84\% (0.0094)   & 99.14\% (0.0040)   & 99.93\% (0.0015)     \\\\[-1em]
Skin\_Seg / 2                & 91.15\% (0.1138)   & 99.93\% (0.0004)   & 99.97\% (0.0003)   & 99.99\% (0.0002)     \\\\[-1em]
\textit{Poker\_Hand / 10}    & 88.53\% (0.0711)   & 94.67\% (0.0505)   & 98.17\% (0.0297)   & 99.31\% (0.0168)     \\\\[-1em]
\textit{SpaceNet / 6}        & 89.12\% (0.0492)   & 94.81\% (0.0384)   & 99.24\% (0.0181)   & 99.95\% (0.0006)     \\ \hline\\[-0.8em]
Average                      & 90.24\%            & 96.08\%            & 99.11\%            & 99.76\%              \\ \hline\\[-1em]
\end{tabularx}
\end{table*}
\begin{figure}
\begin{subfigure}{0.96\linewidth}
    \centering
    \includegraphics[width=\linewidth]{image/chart1.pdf}
    \caption{The k-means algorithm}
    \label{fig:sfig9}
\end{subfigure}
\begin{subfigure}{0.96\linewidth}
    \centering
    \includegraphics[width=\linewidth]{image/chart2.pdf}
    \caption{The EM algorithm}
    \label{fig:sfig10}
\end{subfigure}
\caption{The percentage of computation time with different desired accuracies by using k-means and EM}
\label{chart}
\end{figure}
\begin{figure}
\begin{subfigure}{0.86\linewidth}
    \centering
    \includegraphics[width=\linewidth]{image/box1.pdf}
    \caption{The k-means algorithm}
    \label{fig:sfig11}
\end{subfigure}
\begin{subfigure}{0.91\linewidth}
    \centering
    \includegraphics[width=\linewidth]{image/box2.pdf}
    \caption{The EM algorithm}
    \label{fig:sfig12}
\end{subfigure}
\caption{The box plots of desired accuracy and achieved accuracy}
\label{box}
\end{figure}

\textbf{Achieved Accuracy Validation}. In the experiment, we record the expected stop point for different desired accuracies and calculate the real achieved accuracies. From Table~\ref{kmeans} and Table~\ref{em}, we can see that the average achieved accuracies are 92.09\%, 95.47\%, 98.50\% and 99.81\% when corresponding desired accuracies are 90\%, 95\%, 99\%, 99.9\% respectively for the k-means algorithm. Similarly, for the EM algorithm, the average achieved accuracies are 90.24\%, 96.08\%, 99.11\% and 99.76\% respectively with the same desired accuracy. The numbers in parentheses represent the standard deviations that are generated by different groups for the data set in the clustering process. Fig.~\ref{fig:sfig11} and Fig.~\ref{fig:sfig12} are the box plots of desired accuracy and achieved accuracy in one group of SpaceNet validation set. It clearly shows that the average of actual achieved accuracy is very close to the desired accuracy for both algorithms. The standard deviation is small especially when desired accuracy reaches 99\% and 99.9\%, which proves the high precision of the quadratic polynomial regression in the experiment. When required accuracy is 90\% and 95\%, the achieved accuracy of EM algorithm has larger variation than that of k-means algorithm, which means k-means has better goodness of fit of the regression model than EM algorithm (corresponding to Fig.~\ref{fig:sfig8}). To sum up, k-means is more stable in achieved accuracy than EM in which more anomalies and larger variation can be observed. 

\subsection{Discussion}
From the experiments, we can draw three main conclusions: 1) the higher the desired accuracy, the longer the computation time (i.e., the less the saved time). Users can save much more money with lower but sufficient accuracy (such as 99\%) by using our proposed approach; 2) The performance of cost effectiveness varies with the data sets. It is undoubtedly that our approach can achieve cost effectiveness for different data sets and can be applied in broader fields; 3) The performance of cost effectiveness varies in different clustering techniques. Compared with the k-means algorithm, though the EM algorithm has higher percentages of time-saving with our approach, it normally takes much longer time to converge than k-means as illustrated in Fig.~\ref{accuracy}. Therefore, the actual computation time of k-means algorithm is usually less than the EM algorithm due to its rapid convergence. In real-world applications, different clustering techniques have different application scenarios, it is up to the users to decide which clustering algorithm to adopted. With our approach, they can all achieve cost effectiveness to various degrees. 

For example, in the case studies with the much more efficient k-means algorithm (see Fig.~\ref{accuracy} for efficiency), we find that achieving 99\% accuracy needs 47.71\%-71.14\% computation cost of 100\% accuracy while the less efficient EM algorithm needs 16.69\%-32.04\% of the total computation cost. More specifically, for the SpaceNet data set, the training process for 100 remote sensing images (using the k-means algorithm) took $1,169.46$ seconds and was only computed once. In Section 2.1, we presented the California land use statistics as example for $423,970$ ${km}^2$ land, which need approximately $2.567\times10^7$ partitioned remote sensing images ($438 \times 406$ pixels) of each covering a $16,520.74 m^2$ land. With our approach, the saved computation time is approximately $19,256.73$ hours when the desired accuracy is 99\%. According to the Amazon EC2 pricing \cite{Amazon}, if we run m5.large virtual machine instances, the saved cloud computation cost amounts to \$$4,082.43$ for California with the total computation cost of \$$14,145.63$. Apparently, the training cost (\$0.039) is negligible to the whole computation cost. In real-world applications, the training process is performed once and when it is finished, we can use the regression model many times. For example, we can use the same regression model to conduct the whole United States land use statistics, which will save huge computation cost up to \$$94,687.49$ in each use.

\subsection{Threats to Validity}
In this section, some key threats to the validity will be discussed as follows. 

\textit{Threats to construct validity}. The main threat to the construct validity is the adopted metric to evaluate the accuracy of every intermediate partition during the clustering iterative process. In the paper, we use the Rand Index as the adopted metric. As introduced in Section 3.2, the Rand Index relies on the final partition in the clustering and it is an external clustering index. In most clustering algorithms, the evaluation criteria are divided into internal and external clustering indices. The internal evaluation criterion is to evaluate the goodness of a data partition without prior knowledge from the data sets, which includes Compactness (CP), Separation (SP), Davies-Bouldin Index (DB), Dunn Validity Index (DVI), etc. \cite{fahad2014survey}. And the external evaluation criterion is to assess how accurately a clustering technique partitions the data relative to their correct class labels. In real-world clustering, it is difficult and impractical to retrieve the correct class labels. Thus, the Rand Index is not the usual choice for the real world big data clustering. However, this threat to validity is minimal because our objective is to explore and demonstrate how to stop a clustering process at some point to achieve high cost-effectiveness. The Rand Index can accurately evaluate how close an intermediate partition to the final partition in the training process while internal indices might not be consistently correlated with Rand Index.

\textit{Threats to conclusion validity}. The central threat to the conclusion validity is the reliability of the final partition of the clustering iterative process as the optimal partition. Since the k-means and EM algorithms do not guarantee a global optimum, they attempt to approximate the optimal partition. Therefore, the final partition of the clustering result is not necessarily the optimal partition. So, Fig.~\ref{accuracy} do not necessarily demonstrate how the intermediate partition approaches the real optimal partition. Nevertheless, we are able to consider the final partition adequately reliable for demonstrating the long tail phenomenon in the clustering process because in the optimal situations, the k-means and EM algorithms are likely to take more time and result in a more significant long tail phenomenon. Thus, the threat to the conclusion validity exists but is not significant.

\textit{Threats to external validity}. The main threat to the external validity in our research is the representativeness of the data sets used in the experiments. In the experiments, we used the 3D Road Network, Skin Segmentation, Poker Hand and SpaceNet data sets. All the data sets are real-world data sets. They may have their own characteristics and thus do not comprehensively present all data sets. However, the main features are familiar such as the negative relationship between change rate of the value of the objective function and the clustering accuracy. In the meantime, the high cost-effectiveness and small standard deviation in all given data sets indicate that the threat to the external validity is minimal. 

\textit{Threats to internal validity}. The crucial threat to internal validity is the selection of the regression model. In the experiments, we found that the quadratic polynomial regression model shows the best result of R-squared and SSE for all given data sets. However, it is impossible to exhaust every data set in the real world to ensure if the quadratic polynomial model is the best regression model. For instance, the regression model may be one degree, three-degree polynomial models or even non-polynomial model such as the exponential model in some special data sets. Nevertheless, different types of regression models can also be applied in our approach to achieve cost-effectiveness for the big data clustering in the cloud. Thus, the threat to the internal validity is minimal.

\section{Related Work}
With the development of the pay-as-you-go model, the IT resources are usually provisioned and utilized by cloud computing. Since the majority of advantages offered by cloud computing are built around the flexibility of the pay-as-you-go cost model, cost-effectiveness has become a critical issue in cloud computing filed.
With the improving cloud services from the cloud vendors, many scientists focus on the performance as well as cost-effectiveness of public cloud services. Intensive research work has been made on the cost-effective computation in the cloud environment. A Semi-Elastic Cluster (SEC) computing model \cite{niu2016building} has been proposed for organizations to reserve and dynamically resize a virtual cloud-based cluster. The race-driven results show that such a model has a 61\% percent cost saving than individual users acquiring and managing cloud resources without causing longer average job wait time. And a new MapReduce cloud service model Cura was presented to provide a cost-effective solution to efficiently handle MapReduce production resources, which implemented a globally efficient resource allocation scheme that significantly reduces the resource usage cost in the cloud. A new task scheduler – Flutter \cite{hu2018time}, was designed and implemented which reduces both the completion time and network cost of big data processing jobs across geographically distributed data centers. 

Cost-effectiveness of scientific computing applications has also been studied by Berriman et al. using Amazon’s EC2 \cite{berriman2010application}. They compared Amazon’s EC2 with the Abe high-performance cluster and drew the conclusion that the Amazon EC2 offers better performance and value for processor- and memory-limited applications than for I/O-bound applications. A similar study was conducted by Carlyle’s team to compare the computation cost of high-performance in traditional HPC environments and in Amazon’s EC2 environments, using Purdue University’s HPC “community cluster” program \cite{carlyle2010cost}. Their research showed that an in-house cluster is more cost-effective when the organization having sufficient demand that fully utilizes the cluster or having an IT department capable of sustaining IT infrastructure or having cyber-enabled research as a priority. These features of in-house clusters, in fact confirm the flexibility and cost effectiveness of running computation-intensive applications in the commercial clouds. Wang et al. proposed a stochastic multi-tenant framework for investigating the response time of cloud services as a stochastic metric with a general probability distribution \cite{wang2017optimizing}. In a similar study, by comparing between the scaling out strategies with the scaling up strategies, the performance of Amazon’s cloud services was tested with five benchmark applications and scaling up is found more cost-effective in sustaining heavier workload \cite{hwang2016cloud}. To find the minimum cost of storing and regenerating datasets in multiple clouds, a novel algorithm  was proposed which achieved the best trade-off among computation, storage and bandwidth cost in the cloud \cite{yuan2018algorithm}. Jawad et al. \cite{jawad2018robust} proposed a smart power management system to minimize the operation cost of data center, which coordinates the data center workload, diesel generators,  battery bank,renewable power, real-time trade electricity price and day-ahead power market to reduce consumption cost.

The current research for cloud computing shows the popularity of running computation-intensive applications in the cloud, which describes a general overview about cost effectiveness for big data clustering in the cloud through a comparison between the cloud environment and a traditional cluster environment. 
Additionally, to save cost in the cloud, it is also critical for clustering algorithms to improve their efficiency and to reduce processing time. To deal with the problem, many approaches have been proposed. To optimize the k-means algorithm, how to select k appropriate initial centers is a key issue and there have been many pieces of work on this matter \cite{likas2003global,khan2004cluster,bradley1998refining}. For the EM algorithm, Liu et al. used the parameter expansion to accelerate EM \cite{liu1998parameter}. However, such approaches rarely considered the economic efficiency. Up until now, none of the existing research has considered the k-means or EM algorithm from the cost-effective perspective about cost-effective big data clustering in the cloud. He et al. found the phenomenon that achieving 99\% accuracy of k-means only needs an average of 20\%+ of the total computation time \cite{he2017cost} but they did not offer a solution for terminating the clustering algorithm at an early point with the desired accuracy.

In our research, from a different and important perspective, we take a look at the issue of cost-effectiveness– how to achieve a sufficiently satisfactory accuracy at a relatively small proportion of the total cost of achieving 100\% accuracy by stopping the clustering process at an early point before its completion.

\section{Conclusion and Future Work}
In this research, we proposed a novel approach for cutting the unnecessary ‘long tail’ to achieve cost-effective big data clustering in the cloud. Users can achieve sufficiently satisfactory accuracies at the lowest possible costs by setting their desired accuracies. With our approach, both widely used k-means and EM algorithms show high cost effectiveness in the clustering process. For the k-means algorithm, achieving 99\% accuracy only needs 47.71\%-71.14\% of the computation time for achieving a 100\% accuracy. And for the EM algorithm, achieving a 99\% accuracy needs 16.69\%-32.04\%. By applying our proposed approach, the government will save up to \$$94,687.49$ for the United States land use statistics for each run.

To the best of our knowledge, this is the very first paper to achieve cost effectiveness for big data clustering in the cloud by cutting the unnecessary long tail. This work presents a significant first step toward cost-effective clustering in the cloud. As a contribution, our approach can be easily deployed in various fields which need to clustering big data with limited budget. 

Since k-means and EM algorithms may not be suitable for time-series data and spatiotemporal data, in the future, we plan to investigate the cost effectiveness of the clustering algorithms for those data types. In addition, it is also valuable to explore the relationship between the achieved accuracy and acquired accuracy, and control the margin of error by artificial setting.


%

\ifCLASSOPTIONcompsoc
  \section*{Acknowledgments}
\else
  \section*{Acknowledgment}
\fi

The work is partly funded by the China Scholarship Council.

\ifCLASSOPTIONcaptionsoff
  \newpage
\fi



\bibliographystyle{IEEEtran}
\bibliography{main.bib}

\appendices
\counterwithin{table}{section}
\section{}
The notations used in this paper are shown in Table ~\ref{notation}.
\begin{table}[h]
\caption{Table of notations}
\label{notation}
\begin{tabular}{@{}ll@{}}
\toprule
Notation & Definition                                                                                                  \\ \midrule
$D$      & A given data set to be partitioned                                                                          \\
$x$      & A feature point with dimensions                                                                             \\
$k$      & The number of clusters                                                                                      \\
$n$         & \begin{tabular}[c]{@{}l@{}}The number of iterations in the clustering\\   process\end{tabular}              \\
$c_i,c_j$         & The cluster centers                                                                                         \\
$C$         & The cluster centers set                                                                                     \\
$C_i$         & \begin{tabular}[c]{@{}l@{}}The set of data points that are closer to $c_i$ than\\ to $c_j$  for all $j\neq i$ \end{tabular} \\
$P_i$         & The clustering partition at the $i$th iteration.                                                               \\
$Rand(P_i,P_j)$  & The rand index of two partitions $P_i,P_j$\\
$r_i,r_j$         & The clustering accuracy at the th iteration                                                                 \\
$J$         & The objective function of clustering algorithm                                                              \\
$J_i$         & \begin{tabular}[c]{@{}l@{}} The value of the objective function at the $i$th  \\iteration       \end{tabular}                                              \\
$h_i,h_j$         & \begin{tabular}[c]{@{}l@{}}The change rate of objective function at the \\ $i,j$th iteration\end{tabular}
                                                   \\
$\textit{Time}_\textit{comp}$         & The total computation time                                                                                  \\
$\textit{Time}_\textit{train}$         & The computation time for training                                                                           \\
$\textit{Time}_\textit{actual}$         & \begin{tabular}[c]{@{}l@{}} The computation time when accuracy reaches\\ user’s desired accuracy\end{tabular}                                  \\
$\textit{Time}_\textit{full}$         & \begin{tabular}[c]{@{}l@{}}The computation time when clustering reaches\\ 100\%   accuracy\end{tabular}\\
$\textit{Cost}_\textit{effective}$ & The cost effectiveness percentage
\\ \bottomrule
\end{tabular}
\end{table}

%
\vskip -10pt plus -1fil
\begin{IEEEbiography}[{\includegraphics[width=1in,height=1.25in,clip,keepaspectratio]{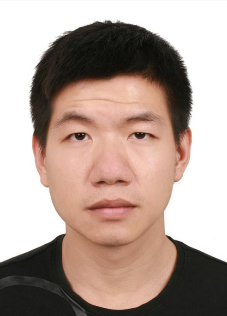}}]{Dongwei Li}
received his M.Sc. degree in software engineering from Wuhan University, China, in 2010. He is currently working toward his Ph.D. degree at Beijing Institute of Technology, Beijing, China. He is a currently visiting researcher at Swinburne University of Technology, Australia. His research interests include data mining and cloud computing.
\end{IEEEbiography}\newpage
\begin{IEEEbiography}[{\includegraphics[width=1in,height=1.25in,clip,keepaspectratio]{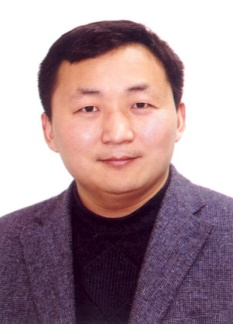}}]{Prof. Shuliang Wang}
received his Ph.D. degree from Wuhan University, China, in 2002 and Hongkong Polytechnic University, Hongkong, in 2003. He is a full professor at Beijing Institute of Technology, China. His research interests include spatial data mining, data field and big data.
\end{IEEEbiography}\vskip -15pt plus -1fil
\begin{IEEEbiography}[{\includegraphics[width=1in,height=1.25in,clip,keepaspectratio]{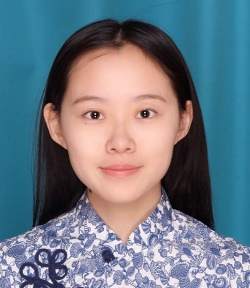}}]{Nan Gao}
received her M.Sc. degree in software engineering from Beijing Institute of Technology, China, in 2018. She is currently working toward her Ph.D. degree at Computer Science and IT, School of Science, Royal Melbourne Institute of Technology, Australia. Her research interests include spatiotemporal data mining and ubiquitous computing.
\end{IEEEbiography}\vskip -16pt plus -1fil
\begin{IEEEbiography}[{\includegraphics[width=1in,height=1.25in,clip,keepaspectratio]{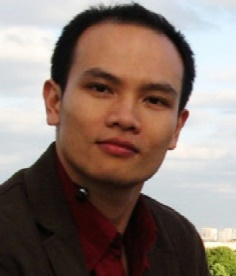}}]{Dr. Qiang He}
received his Ph.D. degree in information and communication technology from Swinburne University of Technology (SUT), Australia, in 2009. He is now a senior lecturer at Swinburne University of Technology. His research interests include software engineering, cloud computing and services computing.
\end{IEEEbiography}\vskip -16pt plus -1fil
\begin{IEEEbiography}[{\includegraphics[width=1in,height=1.25in,clip,keepaspectratio]{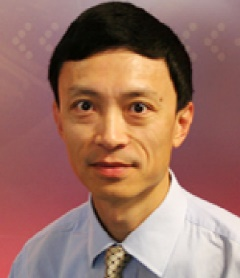}}]{Yun Yang}
received his Ph.D. degree from the University of Queensland, Australia in 1992. He is a full professor at Swinburne University of Technology, Australia. His research interests include distributed systems, cloud computing, software technologies, workflow systems and service-oriented computing. He is the leader of Swinburne’s Next Generation Software Platform focus area and Associate Editor of IEEE Transactions on Cloud Computing.
\end{IEEEbiography}

\newpage




\end{document}